\documentclass[11pt,a4paper,useAMS,usenatbib]{emulateapj}
\usepackage{epsfig}
\usepackage{amsmath}
\usepackage{natbib}
\newcommand\logten{\ensuremath{\log_{10}}}

\begin{document}

\title{Correlation between the total gravitating mass of groups and clusters \\ and the supermassive black hole mass of brightest galaxies}

\author{\'Akos Bogdan\altaffilmark{1}, Lorenzo Lovisari\altaffilmark{1}, Marta Volonteri\altaffilmark{2}, and Yohan Dubois\altaffilmark{2}}
\affil{\altaffilmark{1}Harvard Smithsonian Center for Astrophysics, 60 Garden Street, Cambridge, MA 02138, USA; abogdan@cfa.harvard.edu}
\affil{\altaffilmark{2}Institut d\'Astrophysique de Paris, Sorbonne Universit\'es, UPMC Univ Paris 6 et CNRS, UMR 7095, 98 bis bd Arago, F-75014 Paris, France}
\shorttitle{}
\shortauthors{BOGD\'AN ET AL.}

\begin{abstract}

Supermassive black holes (BHs) residing in the brightest cluster galaxies are over-massive relative to the stellar bulge mass or central stellar velocity dispersion of their host galaxies. As BHs residing at the bottom of the galaxy cluster's potential well may undergo physical processes that are driven by the large-scale characteristics of the galaxy clusters, it is possible that the growth of these BHs is (indirectly) governed by the properties of their host clusters. In this work, we explore the connection between the mass of BHs residing in the brightest group/cluster galaxies (BGGs/BCGs) and the virial temperature, and hence total gravitating mass, of galaxy groups/clusters. To this end, we investigate a sample of 17 BGGs/BCGs with dynamical BH mass measurements and utilize  \textit{XMM-Newton} X-ray observations to measure the virial temperatures and infer the $M_{\rm 500}$ mass of the galaxy groups/clusters. We find that the $M_{\rm BH} - kT$ relation is significantly tighter and exhibits smaller scatter than the $M_{\rm BH} - M_{\rm bulge}$ relations. The best-fitting power-law relations are $ \logten (M_{\rm BH}/10^{9} \ \rm{M_{\odot}})  =  0.20 + 1.74 \logten (kT/1 \ \rm{keV}) $ and $ \logten (M_{\rm BH}/10^{9} \ \rm{M_{\odot}})   =  -0.80 + 1.72 \logten (M_{\rm bulge}/10^{11} \ M_{\odot})$. Thus, the BH mass of BGGs/BCGs may be set by physical processes that are governed by the properties of the host galaxy group/cluster. These results are confronted with the Horizon-AGN simulation, which reproduces the observed relations well, albeit the simulated relations exhibit notably smaller scatter. 
\end{abstract}

\keywords{galaxies: elliptical and lenticular, cD  --- galaxies: evolution ---  galaxies: halos  --- X-rays: galaxies --- X-rays: ISM}

\section{Introduction}
\label{sec:intro}

Observational studies of nearby galaxies demonstrated that galaxies of all morphology types host BHs at their center \citep[for a review, see][]{kormendy13}. By utilizing state-of-the-art telescopes and performing stellar kinematic or gas kinematic measurements of the nuclear regions of galaxies, accurate masses were derived for a sample of nearly 90 BHs. By correlating the BH mass with various properties of the host galaxies, the existence of scaling relations was established. Most notably, BH masses correlate with the velocity dispersion, stellar bulge mass, bulge luminosity, and dark matter halo mass \citep[e.g.][]{magorrian98,gebhardt00,ferrarese02,haring04,gultekin09b,beifiori12,vika12,mcconnell13,lasker14}. The existence of these correlations suggests that the symbiotic growth of BHs and their host stellar bulges and dark matter halos plays a crucial role in galaxy evolution. Specifically, it is hypothesized that during the merger of two (spiral) galaxies, the energetic feedback from BHs can heat and expel the gas from the galaxy. This feedback could quench the active star formation and truncate the rapid growth phase of BHs \citep{silk98,king03,wyithe03,hopkins06}. Alternatively, it may provide positive feedback and trigger star formation at high redshifts \citep{silk05,pipino09}. 

Although there is a general consensus about the importance of BHs in the evolution of galaxies, it remains ambiguous whether the growth of BHs are primarily driven by the stellar bulge or the dark matter halo. Various observational and theoretical studies suggest that -- at least in massive elliptical galaxies -- the primary relation might be between BHs and the host dark matter halos \citep{ferrarese02,bandara09,booth10,bogdan12}. However, the circular velocity, as a proxy for dark matter halo properties, is not a better predictor of BH mass with respect to velocity dispersion  \citep{kormendy11,volonteri11,sun13}. In low-mass,  late-type and bulgeless galaxies, the looser correlation between the BH and dark matter halo may be a consequence of the inefficacy of low-mass galaxies in feeding the BHs, which in turn results in large scatter at the low-mass end of the relation \citep{volonteri11,dubois15,habouzit16}.

\begin{table*}
\caption{Properties of the studied galaxy groups and clusters}
\begin{minipage}{18cm}
\renewcommand{\arraystretch}{1.5}
\centering
\begin{tabular}{c c c c c c c c c c c c}
\hline 
Galaxy & Distance  & $N_{\rm H}$ &$ R_{\rm 500}$& $\rm{R_{frac}}$  & $\sigma$ & $kT$ & $\rm{M_{BH}}$ &      $\rm{\log M_{bulge}}$ & $\rm{M_{500}} $ \\
name &(Mpc)  & ($\rm{10^{20} \ cm^{-2}}$) & (kpc) & & ($\rm{km \ s^{-1}}$) & (keV) & ($\rm{10^{9} \ M_{\odot}}$) & ($\rm{M_{\odot}}$) & ($\rm{10^{13} \ M_{\odot}}$) \\
(1) &(2)  & (3) & (4) & (5) & (6) & (7) & (8) & (9) & (10) \\
\hline

A1836-BCG & 152.4 (a) & $6.21$ &  522 &  1.00 & $288\pm14$ & $1.351\pm0.033$ & $ 3.74^{+0.42}_{-0.52} $ (d) &$ 11.81\pm0.10 $&  $ 4.05\pm0.16$ \\
A3565-BCG & 49.20 (a) & $4.79$ &   454 & 0.40 & $322\pm16$  &   $1.051\pm0.007$ & $ 1.30^{+0.20}_{-0.19}$ (d) &$ 11.78\pm0.09$&  $ 2.68\pm0.03$ \\
IC~1459    & 28.92 (b) & $1.16$ &   363 & 0.24 & $331\pm5$ &   $ 0.696\pm0.01$ & $ 2.48^{+0.48}_{-0.19}$ (e) &$ 11.60\pm0.09 $&  $ 1.36\pm0.03$ \\
NGC~1316   & 20.95 (c)& $2.56$ &   393 & 0.22 & $226\pm9$ &  $0.806\pm0.006$ &  $ 0.169^{+0.038}_{-0.030}$ (f) &$ 11.84\pm0.09 $&  $ 1.73\pm0.02$ \\
NGC~1332   & 22.66 (b)& $2.42$ &   293 &  0.25 & $328\pm9$ & $0.473\pm0.026$ &  $ 1.47^{+0.41}_{-0.20}$ (g) &$ 11.27\pm0.09 $&  $ 0.72\pm0.067$ \\
NGC~1407   & 29.00 (b)& $6.85$ &   459 &  0.19 & $276\pm2$ &  $1.070\pm0.005$ &  $ 4.65^{+0.73}_{-0.41}$ (h)  &$ 11.74\pm0.09 $&  $ 2.76\pm0.25$ \\
NGC~1550   & 52.50 (a)& $16.2$ &   517 &  0.35 & $270\pm10$ &  $1.329\pm0.001$ & $ 3.87^{+0.61}_{-0.71}$  (h) &$ 11.33\pm0.09 $&  $ 3.95\pm0.05$ \\
NGC~3091   & 53.02 (a) &  $4.68$ &  400 & 0.48 & $297\pm12$ &  $0.835\pm0.011$ &  $ 3.72^{+0.11}_{-0.51}$ (h) &$ 11.61\pm0.09 $&  $ 1.83\pm0.04$ \\
NGC~3585   & 20.51 (b) & $6.41$ &   247 & 0.33 & $213\pm11$ & $0.347\pm0.008$ &   $ 0.329^{+0.145}_{-0.058}$ (i) &$ 11.26\pm0.09 $&  $ 0.43\pm0.17$ \\
NGC~3607   & 22.65 (b)& $1.44$ &   306 & 0.15 & $229\pm11$ &  $0.512\pm0.026$ &  $ 0.137^{+0.045}_{-0.047}$ (i) &$ 11.26\pm0.09 $&  $  0.082\pm0.07$ \\
NGC~3842   & 92.20 (a)& $1.67$ &   942 & 0.32 & $270\pm27$ &  $3.963\pm0.065$ &  $ 9.09^{+2.34}_{-2.81}$ (j) &$ 11.77\pm0.09 $&  $ 23.90\pm0.70$ \\
NGC~4291   & 26.58 (b)& $3.26$ &   358 & 0.24 & $242\pm12$ &  $0.682\pm0.002$ &  $ 0.978^{+0.308}_{-0.308}$ (k) &$ 10.85\pm0.09 $&  $ 1.31\pm0.05$ \\
NGC~4486   & 16.68 (c)& $2.11$ &   647 & 0.09 & $324^{+28}_{-12}$ &  $1.997\pm0.001$ &  $ 6.15^{+0.38}_{-0.37}$ (l) &$ 11.72\pm0.09 $&  $ 7.73\pm0.01$ \\
NGC~4889   & 102.0 (a) &  $0.87$ & 1395 & 0.24 & $347\pm5$ & $8.165\pm0.037$  &  $ 20.80^{+15.80}_{-15.90}$  (j) &$ 12.09\pm0.09 $&  $ 78.90\pm0.60$ \\
NGC~6251   & 108.4 (a)& $7.33$ &   334 &  1.00 & $290\pm14$ & $0.598\pm0.049$ &  $ 0.614^{+0.204}_{-0.205} $ (m) &$ 11.88\pm0.09 $& $ 1.06\pm0.15$  \\
NGC~7052   & 70.40 (a)&  $13.8$ &  387 & 0.59 & $266\pm13$ &  $0.785\pm0.023$ &  $ 0.396^{+0.276}_{-0.156}$ (n) &$ 11.61\pm0.10 $&  $ 1.66\pm0.08$ \\ 
NGC~7619   & 53.85 (b)&  $6.26$ &  442 & 0.41 & $292\pm5$ &  $0.997\pm0.005$ &  $ 2.30^{+1.15}_{-0.11}$ (h) &$ 11.65\pm0.09 $&  $ 2.46\pm0.02$ \\
 \hline \\
\end{tabular} 
\end{minipage}
Columns are as follows: (1) Name of the BGG/BCG; (2) Distance to the galaxy. References are: (a) \citet{kormendy13}; (b) \citet{tonry01}; (c) \citet{blakeslee09}; (3) Line-of-sight column density toward the galaxy \citep{willingale13}; (4) $R_{\rm 500}$ radius of the group/cluster; (5) Fraction of the $R_{\rm 500}$ radius that is included within the \textit{XMM-Newton} FOV; (6) Central stellar velocity dispersion of the BGG/BCG taken from \citet{kormendy13}; (7) Best-fit temperature of the hot X-ray emitting gas;  (8) BH mass obtained through dynamical modeling. References are: (d) \citet{dalla09}; (e) \citet{cappellari02}; (f) \citet{nowak08}; (g) \citet{rusli11}; (h) \citet{rusli13}; (i) \citet{gultekin09a}; (j) \citet{mcconnell12}; (k) \citet{schulze11}; (l) \citet{gebhardt11}; (m) \citet{ferrarese99}; (n) \citet{marel98}; (9) Stellar bulge mass taken from \citet{kormendy13}; (10) $M_{\rm 500}$ mass inferred from the best-fit gas temperature given in column (5), and the $kT-M_{\rm 500}$ scaling relation established in \citet{lovisari15}. The errors associated with $M_{\rm 500}$ were computed from the temperature uncertainties. 
\label{tab:list2}
\end{table*}

In this context, it is especially interesting to investigate BHs that reside at the bottom of the potential well of galaxy groups and clusters in BGGs/BCGs. These are among the most massive BHs ever detected, and curiously many of these BHs are over massive for the stellar bulge mass or central stellar velocity dispersion of their host BGG/BCG \citep{gebhardt11,mcconnell11,larrondo12,ferre15,savorgnan15}. The preferential location of these BHs at the bottom of a deep potential well suggests that they may undergo a different evolutionary path than BHs in satellite galaxies. Specifically, centers of groups and clusters may retain a notable amount of gas that could be supplied to the BH after the star-formation has been quenched; cold gas from inflows could be directly funneled onto the center of the group/cluster and directly feed the BHs; or low angular momentum mergers -- that are believed to form the most massive galaxies -- may play a role in increasing the BH mass. Because these processes occur in centers of groups/clusters and likely depend on the large-scale characteristics of the group/cluster, it is sensible to probe whether there is a correlation between the BH mass of BGGs/BCGs and the large-scale properties, such as gas temperature or $M_{\rm 500}$ mass\footnote{The radius $r_{\rm 500}$ is the radius of a sphere where the mean mass density is $500 \rho_{\rm crit}$, where $ \rho_{\rm crit}$ is the critical density of the universe. The mass enclosed within $r_{\rm 500}$ is $M_{\rm 500}$.}, of the host groups and clusters. To probe the existence of such a correlation, we rely on dynamically measured BHs and utilize \textit{XMM-Newton} X-ray observations to characterize the dark matter halo properties.  \\

The paper is structured as follows. In Section 2 we introduce the analyzed sample and our selection criteria. In Section 3 we discuss the analysis of the \textit{XMM-Newton} data. Our results are presented in Section 4, where we derive the best-fit gas temperatures, compute the $M_{\rm 500}$ mass, and investigate the tightness of the $M_{\rm BH} - kT$,  $M_{\rm BH} - M_{\rm 500}$ and $M_{\rm BH} - M_{\rm bulge}$ relations. In Section 5 we place our results in context and compare the observed relations with those obtained from the Horizon-AGN simulations. Or results are summarized in Section 6.  In this work, we assume $H_{\rm{0}}=70 \ \rm{km \ s^{-1} \ Mpc^{-1}}$, $ \Omega_M=0.3$, and $\Omega_{\Lambda}=0.7$, and all error bars represent $1\sigma$ uncertainties.

\section{Sample selection}
\label{sec:sample}

As we aim to explore the connection between the BH mass of BGGs/BCGs and the $M_{\rm 500}$ of galaxy groups/clusters, we require accurate measurements of these quantities. While the $M_{\rm 500}$ of groups and clusters can be derived from relatively short X-ray observations, measuring BH masses using dynamical methods requires state-of-the-art optical telescopes and complex modeling. Therefore, dynamical BH mass measurement are limited to the most massive and/or nearby BHs. As a consequence, only 88 BH masses were measured using dynamical methods until recently \citep{kormendy13}. This set of BHs represents our initial sample. 

From the 88 BHs with dynamical mass measurements, we select those that reside in early-type (E/S0) galaxies, whose systems are typically the BGGs/BCGs of galaxy groups and clusters. This limits our sample to 45 ellipticals and 20 S0 galaxies. Then we utilize the galaxy group and cluster catalog of \citet{tully15} and search for those BHs that reside at the center of galaxy groups/clusters, i.e. are BGG/BCGs. We identified 36 such galaxies, implying that the remaining 29 E/S0 galaxies are satellites. 

To derive the  $M_{\rm 500}$ mass of groups and clusters, we use X-ray scaling relations established between the average gas temperature of the hot intragroup/intracluster medium and the $M_{\rm 500}$ mass of the group/cluster \citep{lovisari15}. To measure the X-ray gas temperatures, we utilize X-ray data taken with the \textit{XMM-Newton} observatory. These observations are ideal to study the hot gas for two reasons. First, the $30\arcmin$ field-of-view (FOV) of \textit{XMM-Newton} allows to explore the galaxy groups/clusters out to a significant fraction of their $r_{\rm 500}$ radius. Second, due to the large collecting area of the \textit{XMM-Newton} cameras, relatively short observations are sufficient to accurately measure the average gas temperature. Therefore, we searched the \textit{XMM-Newton} archive and found that 22 of these groups and clusters have publicly available X-ray observations. 

From this sample, we excluded two bright radio galaxies, Centaurs A and Cygnus A, whose X-ray emission is disrupted by the radio jets, implying that the X-ray scaling relations cannot be applied. From the remaining 20 galaxy groups/clusters we found that the galaxy groups associated with NGC~1194, NGC~2787, and NGC~2960 do not have large-scale diffuse X-ray emission. However, these relatively low stellar mass galaxies ($M_{\rm bulge} \lesssim 10^{11} \ \rm{M_{\odot}}$) reside in the center of poor groups and host low-mass ($M_{\rm BH} <10^{8} \ \rm{M_{\odot}}$) BHs. Therefore, the gravitational potential of these groups may be too shallow to retain a significant amount of hot X-ray gas. In these systems, the faint X-ray emission originates from the optical body of the BGGs. As measuring the X-ray properties of the galaxy gas in these systems would not represent the properties of the large-scale dark matter halo of the groups, we exclude these systems from our analysis. Thus, our final sample consists of 17 groups and clusters. We stress that the BH mass of the sample galaxies were obtained using either stellar or ionized gas dynamics, and our sample does not include any water maser measurements. The properties of the sample are listed in Table \ref{tab:list2}.

\begin{table}
\caption{Exposure times of the analyzed \textit{XMM-Newton} observations}
\begin{minipage}{8.5cm}
\renewcommand{\arraystretch}{1.5}
\centering
\begin{tabular}{c c c}
\hline 
Galaxy &  $t_{\rm total}^{\dagger}$  & $t_{\rm clean}^{\ddagger}$ \\
name & (ks)  & (ks) \\
\hline
A1836-BCG & 36.1/36.2/30.3 & 20.3/22.6/12.4  \\
A3565-BCG & 70.6 70.7 60.9 &  30.6/33.7/19.5 \\
IC 1459    & 131.4/131.9/114.6 & 83.5/82.5/64.6  \\
NGC 1316     & 106.1/106.1/89.4 & 54.7/56.7/28.9 \\
NGC 1332     & 64.7/64.8/56.1 & 49.2/50.1/41.1  \\
NGC 1407     & 64.4/65.3/47.2 & 33.6/34.2/24.5  \\
NGC 1550     & 89.5/87.8/74.1 & 45.5/48.3/29.5 \\
NGC 3091     & 20.9/20.9/16.5 & 18.4/18.7/14.5 \\
NGC 3585     & 20.9/20.7/16.5 & 10.8/10.9/7.3 \\
NGC 3607     & 43.7/43.7/38.1 & 25.1/26.2/17.1  \\
NGC 3842     & 24.2/24.3/16.6 & 21.3/21.5/13.5  \\
NGC 4291     & 50.4/50.4/15.3 & 28.4/32.2/13.2  \\
NGC 4486     & 69.4/68.4/53.9 & 69.4/68.4/53.9 \\
NGC 4889    &  25.0/25.0/18.4   &  18.5/19.2/14.3\\
NGC 6251    &   49.3/49.4/40.9  & 6.8/11.0/2.0 \\
NGC 7052     & 34.2/34.2/28.6 & 15.8/16.6/12.9  \\ 
NGC 7619     & 40.5/40.5/35.4 & 35.6/34.7/28.3  \\
 \hline \\
\end{tabular} 
\end{minipage}
$^{\dagger}$ Total exposure time taken with EPIC MOS1, MOS2, and PN cameras, respectively. \\
$^{\ddagger}$ The clean exposure times refer for the EPIC MOS1, MOS2, and PN cameras, respectively.
\vspace{0.5cm}
\label{tab:list1}
\end{table}

\begin{figure*}[!]
  \begin{center}
    \leavevmode
      \epsfxsize=0.48\textwidth \epsfbox{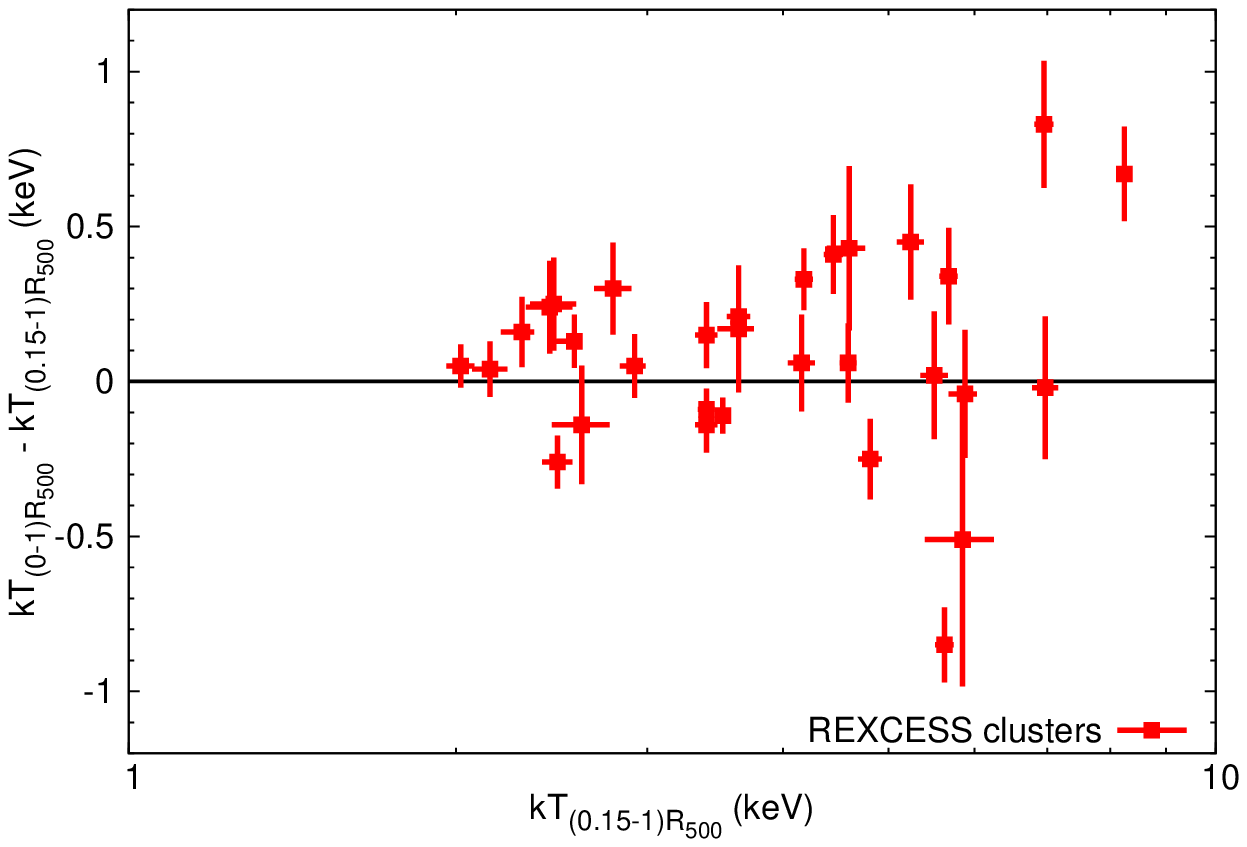}
     \epsfxsize=0.48\textwidth \epsfbox{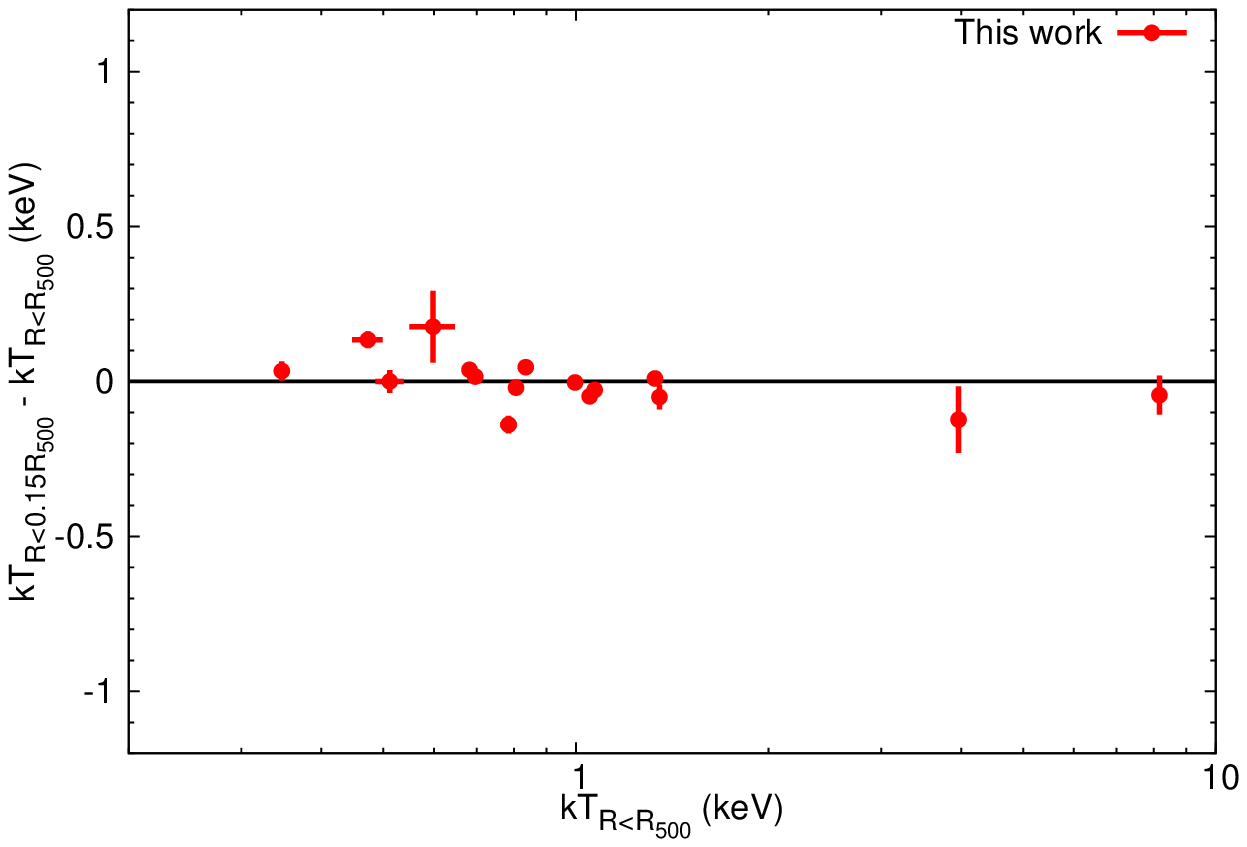}
      \vspace{0cm}
      \caption{Spectroscopic gas temperatures measured in different radial regions. To highlight any offsets, on the y-axis we show the difference between the best-fit gas temperatures. The solid line shows the one-to-one relations. In the left panel, we show the gas temperatures for 31 galaxy clusters from the Representative \textit{XMM-Newton} Cluster Structure Survey \citep{pratt09}.  This relation shows that the best-fit average temperatures are virtually identical if the core region ($R<0.15R_{\rm 500} $) of the clusters is included or excised. In the right panel, we show the best-fit gas temperatures measured for the groups and clusters analyzed in this work. The results -- in agreement with those presented in the left panel -- show that the gas temperatures are nearly identical if they are measured within   $<0.15R_{\rm 500}$ or within a larger region defined by the \textit{XMM-Newton} FOV.} 
     \label{fig:tvar1}
  \end{center}
\end{figure*}

\section{Analysis of the \textit{XMM-Newton} data}
\label{sec:data}

The groups and clusters in our sample were observed with the European Photon Imaging Camera (EPIC) aboard XMM-Newton. Details about the observations are listed in Table  \ref{tab:list2}. To analyze the data, we utilized the XMM Science Analysis System (SAS) version 16.0.0 and Current Calibration Files (CCF). 

We analyzed the data following the main steps described in \citet{lovisari15}. First, we applied \textit{emchain} and \textit{epchain} to the raw data, which tasks generated calibrated event files. For the analysis, we included event patterns 0-12 for EPIC-MOS and 0 for EPIC-PN data. We also removed bright pixels and hot columns and applied out-of-time correction for the EPIC-PN data.

Given that the \textit{XMM-Newton} observations may be severely affected by high background periods, we applied a two-step filtering process. First, we constructed a light curve  using $100$ s binning in the $10-12$ keV and $12-14$ keV bands for EPIC-MOS and EPIC-PN, respectively. By applying $2\sigma$ clipping, we generated a good-time-interval file and excluded the high background periods from the event files. Second, we built light curves using the cleaned event files using $10$ s binning in the $0.3-10$ keV band. Similarly, we applied $2\sigma$ clipping to identify and remove any potentially remaining high background periods. Given that the detector characteristics of EPIC-MOS and EPIC-PN are different, EPIC-PN is significantly more sensitive to high background periods. As a consequence, the clean EPIC-PN exposure time is typically much shorter than that for EPIC-MOS. The resulting total clean exposure time for each analyzed galaxy group and cluster is listed in Table \ref{tab:list1}. 

In this work, we aim to study the diffuse emission associated with galaxy groups and clusters. Therefore, it is essential to identify and exclude bright point sources (mostly background AGN) from the study. To this end, we utilized the task \textit{ewavelet} on the $0.3-10$ keV band images, which produced a list of point sources. Because the source detection algorithm often identifies the luminous cores of groups and clusters as point sources, we checked the source lists by eye and removed these detections. The thus obtained source regions were masked from the further analysis of the diffuse emission. 

An essential part of the data analysis is to precisely account for the particle and cosmic X-ray background (CXB) components. While the influence of background is less pronounced in luminous galaxy clusters, a predominant fraction of our sample consists of relatively faint galaxy groups. Moreover, all systems in our sample fill the entire FOV of the \textit{XMM-Newton} cameras, which further complicates the accurate background subtraction. To account for the quiescent and non-vignetted particle background components, we utilized the filter-wheel closed observations for the EPIC-MOS and EPIC-PN data \citep{freyberg06,snowden08}. These data were re-normalized using the count rates observed in the $3-10$ keV energy range \citep{zhang09}. This re-normalized filter-wheel closed observation was used to subtract the instrumental background components. The emission originating from the cosmic X-ray background was included in the spectral fits. Specifically, we used the spectra from the ROSAT All Sky Survey data that were obtained through the HEASARC webpage\footnote{http://heasarc.gsfc.nasa.gov/cgi-bin/Tools/xraybg/xraybg.pl}. The CXB spectra were extracted from  regions beyond the virial radius of the groups/clusters; hence, in these regions, we do not expect significant emission associated with these systems. We describe the spectrum of the CXB with a three-component model including an absorbed $\sim0.2$ keV thermal model, an unabsorbed $\sim0.1$ keV thermal component, and a power-law model with a slope of $\Gamma=1.41$. The temperature of the thermal models describing the CXB were allowed to vary, but their metallicity was fixed at $1.0$ solar. This model for the CXB was simultaneously fit with the source spectra extracted from the three cameras. For the fitting, the temperature and the metallicities of the hot group/cluster gas were allowed to vary, but they were linked for the three EPIC cameras.

\begin{figure*}[!]
  \begin{center}
    \leavevmode
      \epsfxsize=0.85\textwidth\epsfbox{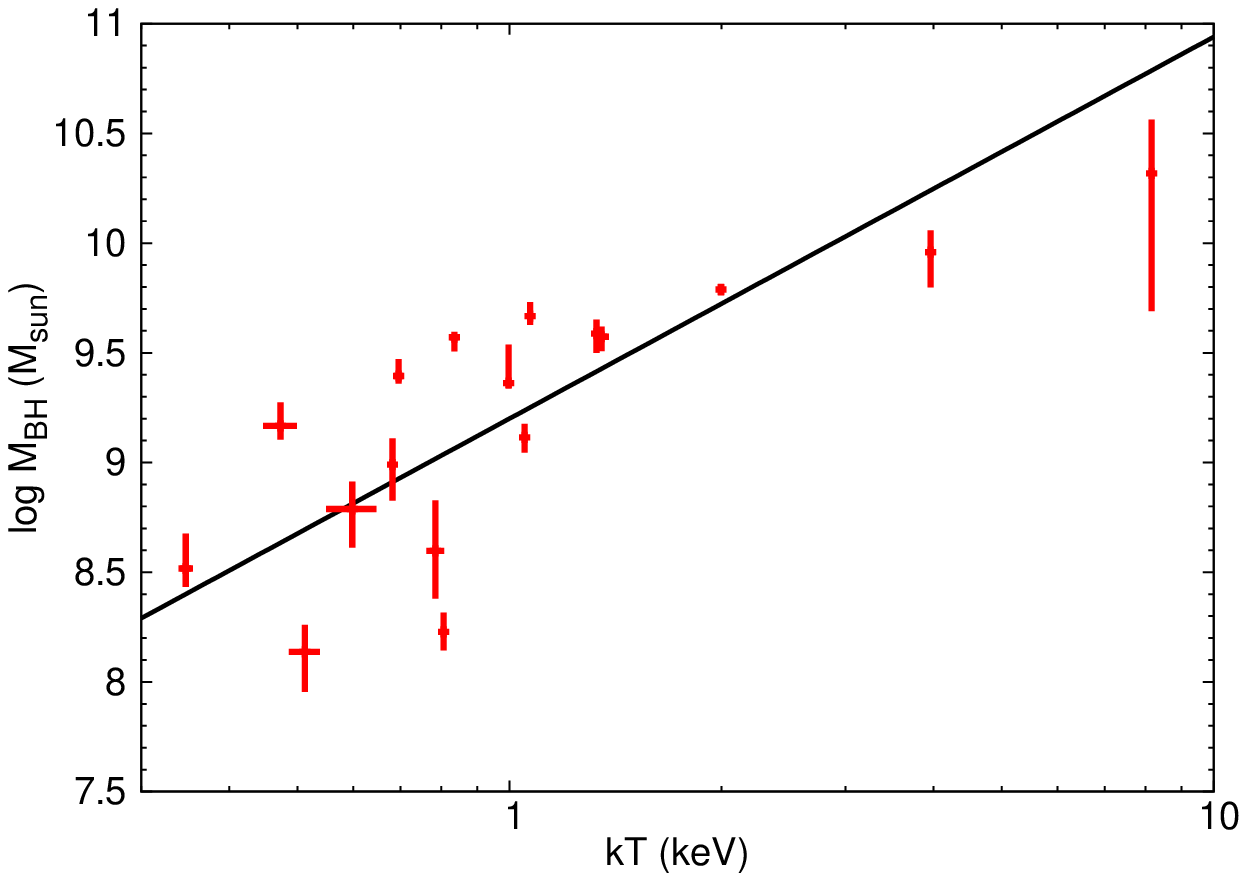}
      \vspace{0cm}
      \caption{Correlation between the BH mass and the best-fit gas temperature of the group/clusters. Note that the gas temperature of the intragroup/intracluster medium is a direct tracer of the total gravitation mass \citep{eckmiller11,lovisari15}. Our sample covers a broad range of systems, including low-mass groups with sub-keV temperatures and massive, $\sim8$ keV, galaxy clusters.  The solid line shows the best-fit power-law relation that can be described as $ \logten (M_{\rm BH}/10^{9} \ \rm{M_{\odot}})  =  0.20 + 1.74 \logten (kT/1 \ \rm{keV}) $. The Pearson and Spearman correlation coefficients of the relation are 0.97 and 0.83, respectively, showing a strong correlation.} 
     \label{fig:mbh_kt}
  \end{center}
\end{figure*}

\begin{figure*}[!]
  \begin{center}
    \leavevmode
      \epsfxsize=0.85\textwidth\epsfbox{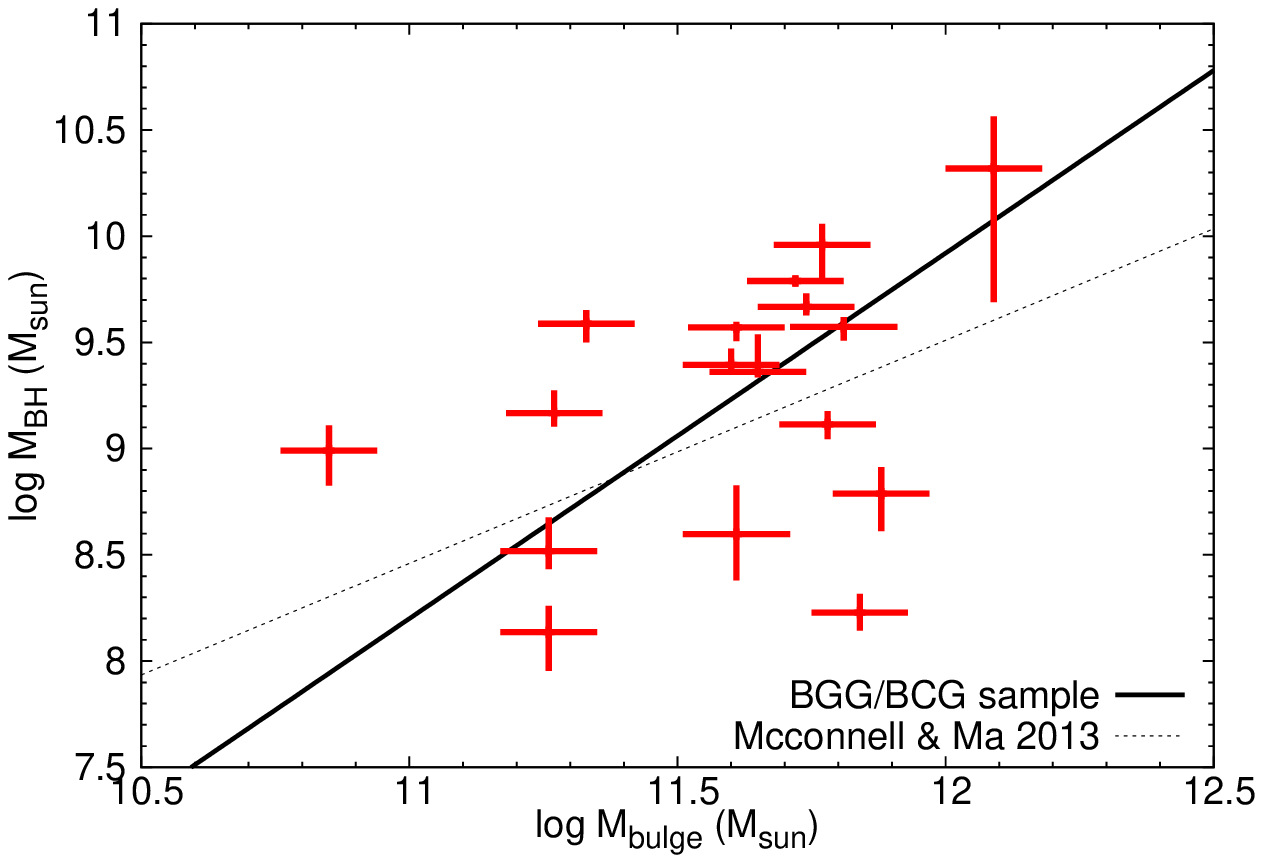}
      \vspace{0cm}
      \caption{Correlation between the BH mass and the stellar bulge mass of BGGs/BCGs. Thick solid line shows the best-fit power-law relation to the 17 data points studied in this work. The best-fit relation is $ \logten (M_{\rm BH} / 10^{9} \ \rm{M_{\odot}}) =  -0.80 + 1.72 \logten (M_{\rm bulge}/10^{11} \ M_{\odot})$. The dashed line shows the best-fit relation obtained for a sample of 72 galaxies with dynamical BH masses \citep{mcconnell13}. Most BGGs/BCGs host over-massive BHs relative to this relation. We note that all 17 BGGs/BCGs that are studied in the current work were included in the study of \citet{mcconnell13}.  The Pearson and Spearman correlation coefficients of the best-fit relation are 0.70 and 0.35, respectively, showing a moderate/weak correlation.} 
     \label{fig:mbh_mbulge}
  \end{center}
\end{figure*}

\section{Results}
\label{sec:results}

\subsection{Measuring the gas temperatures}
\label{sec:temperature}

To probe the total gravitating mass of the galaxy groups and clusters in our sample, we rely on \textit{XMM-Newton} X-ray observations of the hot gas.  Although it would be desirable to compute the total gravitating mass of the groups/clusters by assuming that the hot X-ray gas is in hydrostatic equilibrium, this would require accurate measurements of the density and temperature profiles, preferably out to $R_{\rm 500}$. The relatively X-ray faint nature of several groups in our sample, the short exposure times of the \textit{XMM-Newton} observations, and the \textit{XMM-Newton} FOV that is smaller than the actual $R_{\rm 500}$ make it unfeasible to build temperature profiles for most systems in our sample. However, under the assumption of hydrostatic equilibrium, the virial temperature of the intragroup or intracluster medium is directly proportional with the total mass of the system  \citep{pratt09,eckmiller11,lovisari15}. Thus, we rely on the temperature of the intragroup/intracluster medium to trace the total mass of the groups/clusters.

Because all groups/clusters in our sample are relatively nearby ($D\lesssim150$ Mpc), their $r_{500}$ radius is comparable or exceeds the \textit{XMM-Newton} FOV. Therefore, we extract spectra from circular regions centered on the center of the groups/clusters, whose radii approximately fill the FOV. Given the different virial radii and the different distance of the groups/clusters, the extraction regions vary in the range of $(0.09-1)R_{\rm 500}$ (see Table \ref{tab:list2}). 

For each system, we fit the X-ray spectra with an optically-thin thermal plasma emission model (\textsc{apec} model in \textsc{XSpec}). The metal abundances were free to vary \citep{asplund09} and the column density was fixed at the total column density that includes both the atomic and molecular hydrogen  \citep{willingale13}. Because we extract three spectra from the EPIC cameras, we fit these spectra simultaneously.  For most systems in our sample, the X-ray emission from the intragroup/intracluster medium dominates the overall emission. However, in a few fainter galaxy groups, the emission from the central galaxy may add a noticeable contribution, particularly at energies exceeding $2$ keV. In these cases, we added a power-law model to account for the emission originating from the population of unresolved X-ray binaries \citep{gilfanov04} that are associated with the BGG. For the power-law model, we fixed the slope at $\Gamma=1.56$, which is the average slope of low-mass X-ray binaries \citep{irwin03}. We note that adding a power-law component did not change the best-fit gas temperature, but it improved the overall goodness of the fits. The best-fit gas temperatures are listed in Table \ref{tab:list2}.

\begin{table}
\caption{Best-fit gas temperatures extracted from various regions}
\begin{minipage}{8.5cm}
\renewcommand{\arraystretch}{1.5}
\centering
\begin{tabular}{c c c}
\hline 
Galaxy &  $kT \  ({\rm R\leq R_{500}})$  & $kT \  ({\rm R<0.15R_{500}})$ \\
name & (keV)  & (keV) \\
\hline
1836-BCG   &     $1.351\pm0.033$ & $1.301\pm0.022$ \\ 
A3565-BCG &   $1.051\pm0.007$ & $1.003\pm0.007$ \\
IC~1459      &      $ 0.696\pm0.01$ & $ 0.712\pm0.007$ \\
NGC~1316  &   $0.806\pm0.006$ & $0.786\pm0.004$ \\
NGC~1332  &  $0.473\pm0.026$ & $0.608 \pm 0.008$ \\
NGC~1407  &  $1.070\pm0.005$ & $1.043\pm0.005$ \\
NGC~1550  &  $1.329\pm0.001$ & $1.339 \pm0.002$ \\
NGC~3091  &  $0.835\pm0.011$ & $0.881 \pm0.008$ \\
NGC~3585   & $0.347\pm0.008$ & $ 0.381 \pm0.030$ \\
NGC~3607$^\dagger$   &   $0.512\pm0.026$ &$-$ \\
NGC~3842  &   $3.963\pm0.065$ & $3.840\pm0.085$ \\
NGC~4291  &   $0.682\pm0.002$ & $0.720\pm0.009$ \\
NGC~4486$^\dagger$  &   $1.997\pm0.001$ & $-$ \\ 
NGC~4889  &   $8.165\pm0.037$ & $8.121\pm0.051$ \\
NGC~6251  &  $0.598\pm0.049$ &  $0.775\pm0.105$ \\
NGC~7052  &   $0.785\pm0.023$ &  $0.646\pm0.016$ \\
NGC~7619  &   $0.997\pm0.005$ & $0.994\pm0.004$  \\
 \hline \\
\end{tabular} 
\end{minipage}
$\dagger$ Note that  NGC~3607  and NGC~4486 can only be explored to $0.15R_{\rm 500}$ and  $0.09R_{\rm 500}$, respectively. Therefore, it is not feasible to perform the comparison for these two objects.
\vspace{0.5cm}
\label{tab:temp}
\end{table}

\subsection{Temperature dependence as a function of extraction radius}
\label{sec:radius}

In this work, we use the $kT-M_{\rm 500}$ scaling relation to derive the $M_{500}$ of the groups/clusters in our sample. However, the average gas temperatures presented in   \citet{lovisari15} were obtained from the cool-core excised spectra and refer to the radial region of $(0.15-1)R_{\rm 500}$. The different extraction region applied by \citet{lovisari15} has two main origins. First, the groups in their sample were  -- on average -- more distant; hence, a larger fraction of them could fit in the \textit{XMM-Newton} FOV. Second, they excised the central regions of the groups to exclude any potential cool cores. Given the proximity of the groups/clusters in our sample and the \textit{XMM-Newton} FOV, it is not feasible to probe the characteristics of the X-ray gas out to $R_{\rm 500}$. To map the intragroup/intracluster medium of these systems out to $R_{\rm 500}$, extensive mosaic X-ray observations would be required. However, such  observations are not available in the archive; therefore, we must restrict our study to the regions that are included within the \textit{XMM-Newton} FOV. In addition, for NGC~3607 and NGC~4486 only the innermost regions ($R<0.15R_{\rm 500}$) are included within the FOV (Table \ref{tab:list2}).  Given that we derive the $M_{\rm 500}$ masses from the best-fit gas temperature, we must investigate whether the temperatures  exhibit a notable variation if different extraction regions are used. To probe the temperature variations as a function of radius, we use multiple approaches.

First, we investigate the temperature profiles of the galaxy groups studied in \citet{lovisari15}. Those temperature profiles, which can be approximated with the self-similar temperature profile of \citet{vikhlinin06}, exhibit the peak temperatures at $\sim0.1 R_{\rm 500}$ and level off at larger radii. For most systems the temperature difference between the peak temperature and the lowest temperatures reached in the outskirts is $\lesssim30\%$. This value should be considered as an upper limit on the temperature uncertainty, as the best-fit temperatures for our sample galaxies are luminosity-weighted values computed from regions, which include the peak and part of the flattening profile.  

Second, we utilize the results of \citet{pratt09}, who investigated 31 galaxy clusters from the Representative \textit{XMM-Newton} Cluster Structure Survey (REXCESS). These authors measured the gas temperature in three different regions: $(0-1)R_{\rm 500}$,  $(0.15-1)R_{\rm500}$, and $(0.15-0.75)R_{\rm500}$. In the left panel of Figure \ref{fig:tvar1} we present the scatter plot showing the best-fit temperatures obtained between $R<R_{\rm500}$ and $(0.15-1)R_{\rm500}$. This plot demonstrates that the average gas temperatures are virtually identical whether they are measured at radii $R<0.15R_{\rm500}$ or $(0.15-1)R_{\rm500}$. Thus, based on the results obtained for the REXCESS sample, it appears unlikely that including or excluding the innermost regions of groups/clusters will significantly influence the overall temperatures. To further confirm this, we compared the gas temperatures for two different regions for the groups and clusters in our sample. Namely, we extracted spectra from  $R<0.15R_{\rm500}$ and from the full apertures.  In the right panel of Figure \ref{fig:tvar1} we show the scatter plot obtained between these two temperatures and we tabulate the best-fit values in Table \ref{tab:temp}. This reveals that the temperatures are nearly uniform and do not show a notable variation whether the innermost $R<0.15R_{\rm500}$ region is excised. 

Third, we investigated six galaxy groups/clusters from our sample, which are sufficiently distant and allow us to extract spectra and measure temperatures out to $\geq0.4R_{\rm 500}$. Specifically, we selected Abell~1836, Abell~3565, NGC~3091, NGC~7052, NGC~6251, and NGC~7619. For each of these systems, we extracted spectra within the radii $R<0.1R_{\rm 500}$, $R<0.2R_{\rm 500}$, $R<0.3R_{\rm 500}$, and $R<0.4R_{\rm 500}$. We fit the spectra following the method described in Section \ref{sec:temperature} and measured the gas temperatures in each of these regions. The results point out that the temperatures measured within these radii agree within $10\%$.  

Thus, we conclude that measuring the temperatures of the intragroup/intracluster medium within radii that are primarily defined by the FOV will not influence our conclusions in any significant way. Specifically, it may add a systematic uncertainty of no more than $10\%$. This uncertainty in the gas temperature corresponds to a $\leq20\%$ systematic uncertainty in the inferred $M_{\rm 500}$ values.

\begin{table*}
\caption{Best-fit parameters on linear regression}
\begin{minipage}{18cm}
\renewcommand{\arraystretch}{1.5}
\centering
\begin{tabular}{c c c c c c c}
\hline 
Relation &  $\alpha$  & $\beta$ & $\sigma_X$ & $\sigma_Y$ & $r^\dagger$ & $\rho^\ddagger$ \\
\hline
$M_{\rm BH} - kT$ & $0.20 \pm 0.09$ & $1.74  \pm0.16$    & 0.22 & 0.38 & 0.97 & 0.83\\
$M_{\rm BH} - M_{\rm 500}$ & $-0.22 \pm0.13$ &  $1.07\pm0.17$  & 0.36 & 0.38 & 0.97 & 0.83\\
$M_{\rm BH} - M_{\rm bulge}$ & $-0.80\pm0.23$ &  $1.72\pm0.27$  & 0.35 & 0.61 & 0.70 & 0.35\\
$M_{\rm BH} - \sigma$ & $0.51\pm0.11$ &  $9.68\pm0.15$  & 0.05 & 0.46 & 0.51 & 0.58 \\
$M_{\rm bulge} - kT$ & $ 0.58\pm0.07$ &  $0.87\pm0.24$ & 0.30 & 0.26 &0.73 & 0.60  \\
 \hline
$M_{\rm BH} - M_{\rm 500}$ (simulated) & $-0.05 \pm0.01$ &  $0.92 \pm0.06$  & 0.20 & 0.19 & 0.82 & 0.84 \\
\end{tabular} 
\end{minipage}
$^\dagger$ Pearson correlation coefficient \\
$^\ddagger$ Spearman correlation coefficient
\vspace{0.5cm}
\label{tab:fit}
\end{table*}

\subsection{Measuring $M_{\rm 500}$}
\label{sec:halomass}

Based on the obtained gas temperatures, we infer  $M_{\rm 500}$ by utilizing the $kT-M_{\rm 500}$ scaling relation from \citet{lovisari15}: 

$$ M_{\rm 500} = 7.744\times10^{13}   (kT/2 \ \rm{ keV})^{1.65} \ \rm{M_{\odot}} \ . $$

The slope and normalization of this relation is in good agreement with that obtained for more massive clusters; hence, the  \citet{lovisari15} relation is essentially the extension of the $kT-M_{\rm 500}$ relation established for clusters  \citep[e.g.][]{arnaud05}. The results presented in \citet{lovisari15} demonstrate that the $kT-M_{\rm 500}$ relation can be extended to the lower-mass end of groups. Thus, the gas temperatures of the groups/clusters obtained in this work can be robustly used to determine their total gravitating mass. 

Based on the best-fit temperatures, we obtained $M_{\rm 500}$ masses in the range of $(0.43-78.90)\times10^{13} \ \rm{M_{\odot}}$; hence, our sample groups and clusters cover more than two orders of magnitude in $M_{\rm 500}$. The values for each system are given in tabulated form in Table \ref{tab:list2}.

\begin{figure}[!]
  \begin{center}
    \leavevmode
      \epsfxsize=0.48\textwidth\epsfbox{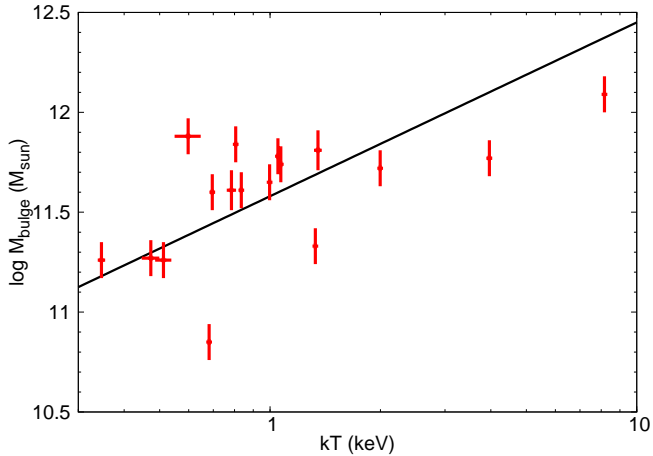}
      \vspace{0cm}
      \caption{Stellar bulge mass as a function of gas temperature for a sample of 17 systems studied in this work. The solid line shows the best-fit relation that can be described as $ \logten ({M_{\rm bulge}  / 10^{9} \ \rm{M_{\odot}}  })  =  0.58 + 0.87 \logten (kT/1 \ \rm{keV}) $. The Pearson and Spearman correlation coefficients of the relation are 0.73 and 0.60,  showing a moderate correlation.} 
     \label{fig:kt_mbulge}
  \end{center}
\end{figure}

\begin{figure}[!]
  \begin{center}
    \leavevmode
      \epsfxsize=0.48\textwidth\epsfbox{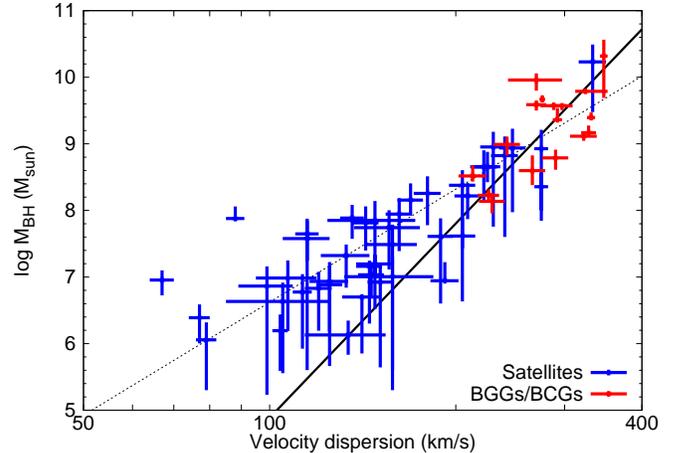}
      \vspace{0cm}
      \caption{BH mass as a function of velocity dispersion for the 17 BGGs/BCGs in our sample and all galaxies with dynamical BH mass measurements \citep{kormendy13}. The thick solid line shows the best-fit relation for the BGGs/BCGs, whereas the dashed line shows the best-fit relation from  \citet{mcconnell13}. Note that the BGG/BCG sample produces a steeper relation than that obtained for the entire BH sample. } 
     \label{fig:mbh_sigma}
  \end{center}
\end{figure}

\begin{figure*}[!]
  \begin{center}
    \leavevmode
      \epsfxsize=0.48\textwidth\epsfbox{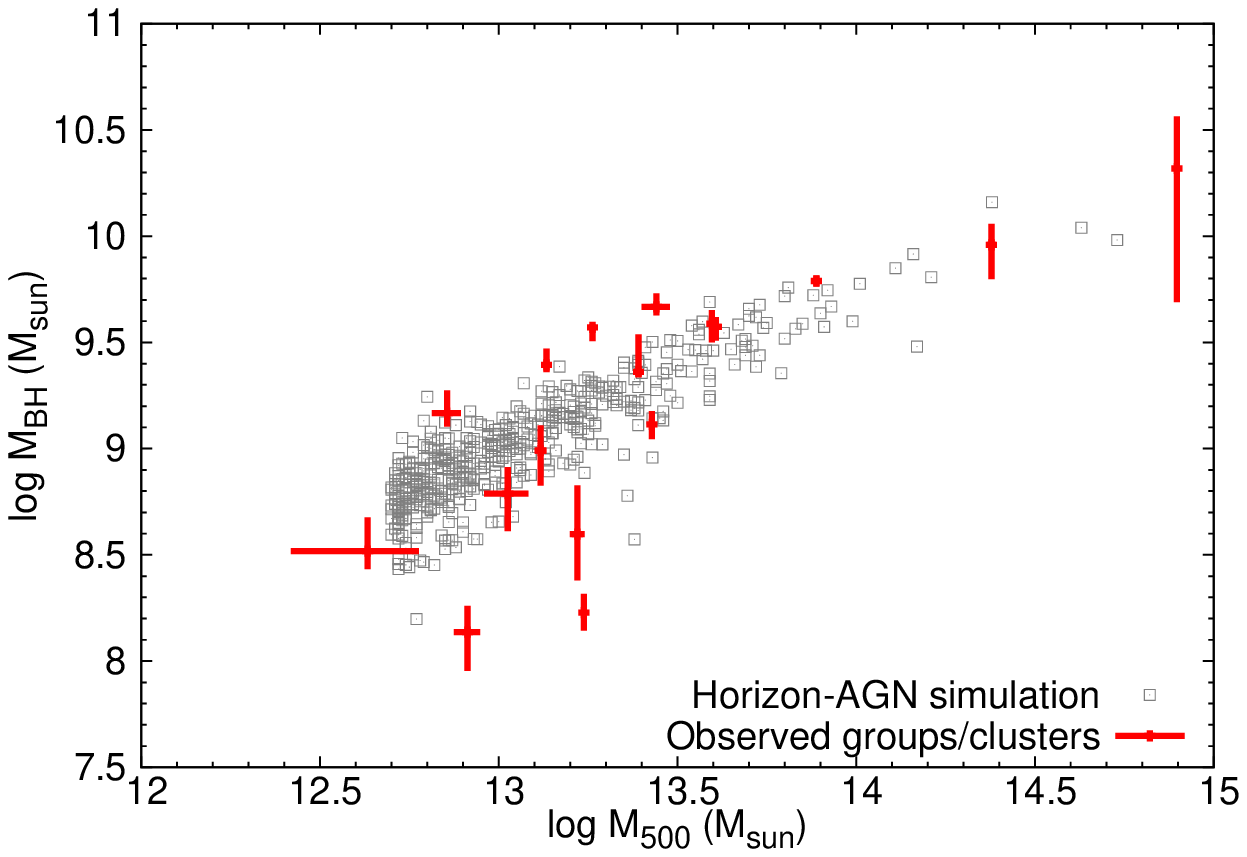}
      \epsfxsize=0.48\textwidth\epsfbox{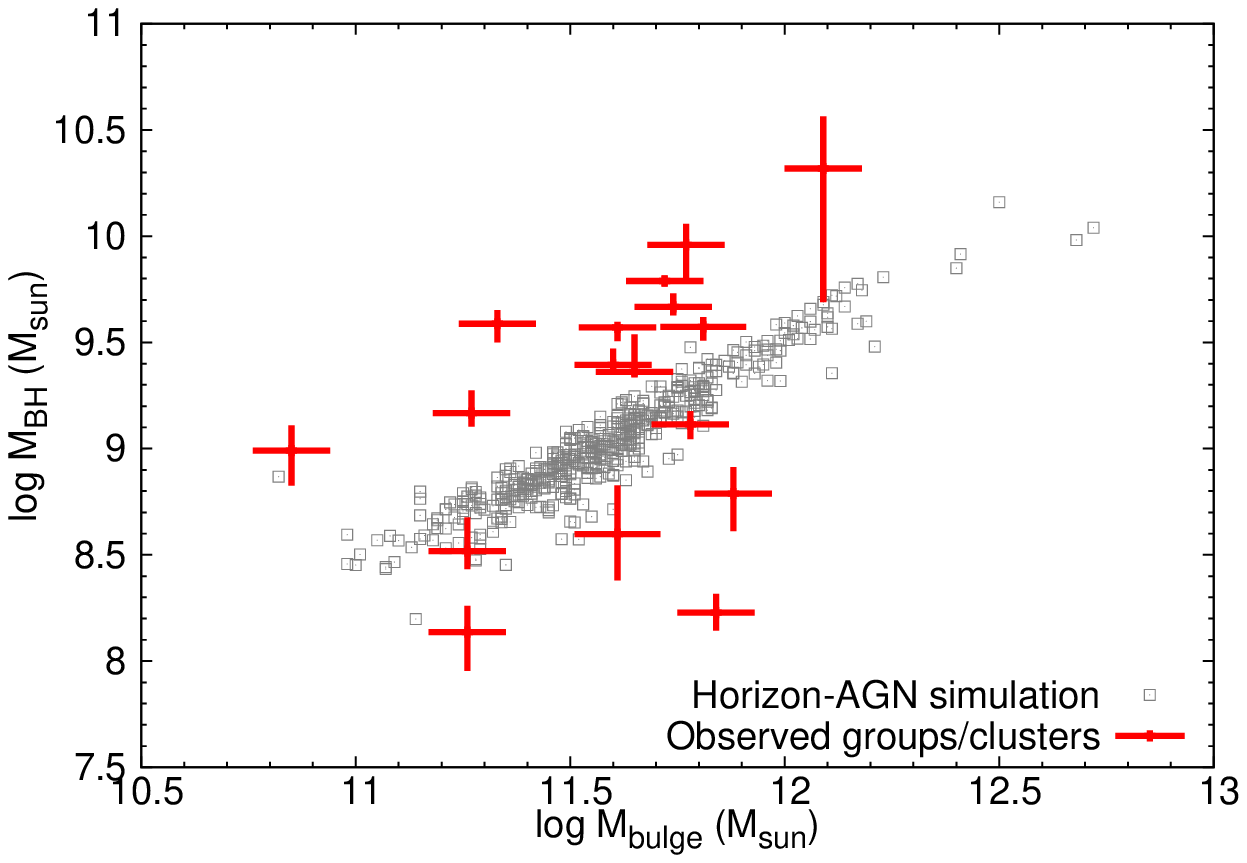}
      \vspace{0cm}
      \caption{Left: BH mass as a function of $M_{\rm 500}$ (left panel) and $M_{\rm bulge}$ (right panel). In these panels, the observed data for the sample of 17 galaxy groups/clusters is compared with the groups/clusters from the the Horizon-AGN simulations. There is an overall good agreement between the observed and simulated relations. However, the simulated relations, especially the $M_{\rm BH}  - M_{\rm bulge}$ exhibit smaller scatter than the observed ones. } 
     \label{fig:fig1}
  \end{center}
\end{figure*}

\subsection{Correlations}
\label{sec:correlations}

Based on the dynamical BH mass measurements (see Table \ref{tab:list2}), the stellar bulge masses \citep[taken from][]{kormendy13}, and the best-fit temperatures measured for the groups/clusters, we investigate the following three relations: $M_{\rm BH} - kT$, $M_{\rm BH} - M_{\rm bulge}$,  $M_{\rm bulge} - kT$.  We fit each of these relations in log space using the functional form of: 
$$ \logten Y =  \alpha + \beta \logten X \ \rm{,}$$
where $\alpha$ and $\beta$ represents the intercept and the slope, respectively, and $X$ and $Y$ are given in units of $M_{\rm BH}/(10^{11} \ M_{\odot})$, $kT/(1 \ \rm{keV}) $, and $M_{\rm bulge}/(10^{11} \ \rm{M_{\odot}}$). 

We used the BCES\_REGRESS code to perform the linear regression and to derive the intrinsic scatter in the relations \citep{akritas96}. We computed the fits in log-log space using the bisector fitting method of  the BCES code.  The logarithm scatter,  $\sigma_X$  and $\sigma_Y$, along the $X$- and $Y$-axis were computed following the formalism described in \citet{lovisari15}. The best-fit parameters and the scatter of the fits are listed in Table \ref{tab:fit}. The obtained scaling relations are the following: 

$$\log_{10} \Big( \frac{M_{\rm BH}}{10^{9} \ \rm{M_{\odot}}} \Big) = 0.20 + 1.74 \log_{10}  \Big( \frac{kT}{1  \ \rm {keV}}\Big) $$ 

$$\log_{10} \Big( \frac{M_{\rm BH}}{10^{9} \ \rm{M_{\odot}}} \Big) =  -0.22 + 1.07 \log_{10}  \Big( \frac{M_{\rm 500}}{10^{13}  \ \rm {M_{\odot}}}\Big) $$ 

$$\log_{10} \Big( \frac{M_{\rm BH}}{10^{9} \ \rm{M_{\odot}}} \Big) = -0.80 + 1.72 \log_{10}  \Big( \frac{M_{\rm bulge}}{10^{11}  \ \rm {M_{\odot}}}\Big) $$ 

$$\log_{10} \Big( \frac{M_{\rm BH}}{10^{9} \ \rm{M_{\odot}}} \Big) = 0.51 + 9.68 \log_{10}  \Big( \frac{\sigma}{300  \ \rm {km \ s^{-1}}}\Big) $$ 
	
$$\log_{10} \Big( \frac{M_{\rm bulge}}{10^{11}  \ \rm {M_{\odot}}}\Big) = 0.58 + 0.87 \log_{10}  \Big(\frac{kT}{1  \ \rm {keV}} \Big) $$ 

We depict the observed data points with the best-fit relations in Figures \ref{fig:mbh_kt}, \ref{fig:mbh_mbulge}, and \ref{fig:kt_mbulge}. 
 
The results of the linear regression point out that the $M_{\rm BH} - kT$ relation (Figure \ref{fig:mbh_kt}) is significantly tighter than the $M_{\rm BH} - M_{\rm bulge}$ relation (Figure \ref{fig:mbh_mbulge}). In addition, we probe the tightness of the relations by computing the Pearson and Spearman correlation coefficients. These point out a strong correlation for  the $M_{\rm BH} - kT$ relation ($r=0.95$, $\rho=0.83$) and a weak correlation for the  $M_{\rm BH} - M_{\rm bulge}$ relation ($r=0.70$, $\rho=0.35$). Thus, in the sample of 17 galaxy groups/clusters the BH mass traces the gas temperature significantly more tightly than the stellar bulge mass. 

We also probe $M_{\rm bulge} - kT$ to test whether the stellar bulge mass and the gas temperature of the groups/clusters exhibit a tight correlation (Figure \ref{fig:kt_mbulge}). We find that the scatter of this relation exceeds that of the $M_{\rm BH} - kT$ relation, and the Pearson and Spearman correlation coefficients denote a moderate correlation (Table \ref{tab:fit}). This suggests that the tight correlation between the BH mass and group/cluster temperature is not the consequence of -- an even tighter -- relation between the stellar bulge mass and the group/cluster temperature.

In Figure \ref{fig:mbh_sigma}, we show the $M_{\rm BH} - \sigma$ relation for the 17 BGGs/BCGs and extend the sample with galaxies with dynamical BH masses. To this end, we utilized the BH catalog compiled by \citet{kormendy13}, which includes additional 69 galaxies. We find that the best-fit $M_{\rm BH} - \sigma$  relation  exhibits a significantly steeper slope than the relation established for the full galaxy sample.

\section{Discussion}

\subsection{The Horizon-AGN simulation}
\label{sec:simulation}

We compare our results to the Horizon-AGN simulation \citep{dubois14}. This simulation is run with the Adaptive Mesh Refinement code {\sc ramses}~\citep{teyssier02} and includes prescriptions for background UV heating, gas cooling including the contribution from metals released by stellar feedback, star formation following a Schmidt law with a 1 per cent efficiency~\citep{rasera&teyssier06}, and feedback from stellar winds and type Ia and type II supernovae (SNae) assuming a Salpeter initial mass function \citep{dubois&teyssier08,kaviraj17}. BH formation, accretion, and feedback are included in the simulation. The accretion onto BHs follows the Bondi-Hoyle-Lyttleton rate, capped at the Eddington luminosity with an assumed radiative efficiency of $\epsilon_{\rm r}=0.1$. AGN feedback is a combination of two different modes: for $\dot M_{\rm BH}/\dot M_{\rm Edd}>0.01$, an isotropic injection of thermal energy into the gas within a sphere of radius $\Delta x=1$ proper kpc (corresponding to the simulation effective resolution), at an energy deposition rate: $\dot E_{\rm AGN}=\epsilon_{\rm f} \epsilon_{\rm r} \dot M_{\rm BH}c^2$, where $\epsilon_{\rm f}=0.15$ was chosen to reproduce the correlations between BHs and galaxies and the BH density in our local universe (see \citealp{2012MNRAS.420.2662D}). At low accretion rates, $\dot M_{\rm BH}/\dot M_{\rm Edd}<0.01$,  AGN feedback energy, with $\epsilon_{\rm f}=1$, is deposited into a bipolar outflow with a jet velocity of $10^4\,\rm km\, s^{-1}$ into a cylinder with a cross section of radius $\Delta x$ and height $2 \, \Delta x$ following~\citep[][]{ommaetal04}. Further details about the jet implementation are given in~\citealp{duboisetal10}.

We identify dark matter halos and sub-halos using HaloMaker, which uses AdaptaHOP \citep{Aubert+04,Tweed+09}, a structure-finder based on the identification of saddle points in the (smoothed) density field, using the shrinking sphere approach proposed by~\cite{poweretal03} to determine the halo center.  Only dark matter halos identified with more than 50 particles are considered. Galaxies are identified using the same method and same parameters but using the stellar particle distribution instead of the dark matter one. We assign a BH to a halo if it is within 10 per cent of the virial radius of the dark matter halo and to a halo+galaxy structure if the BH is also within twice the effective radius of the most massive galaxy within that halo. If more than one BH meets the criteria, the most massive BH is defined as the central BH \citep{volonteri2016}. 

The size of the box is $L_{\rm box}=100 \, h^{-1} {\rm Mpc}$, and we identified 484 groups and clusters with $M_{\rm 500}$ larger than $5\times10^{12} \ \rm{M_{\odot}}$ at $z=0$. In this paper, we analyze the central galaxy and respective BH for each group and cluster. We define the bulge mass as the total stellar mass.

\begin{figure}[!]
  \begin{center}
    \leavevmode
      \epsfxsize=0.48\textwidth\epsfbox{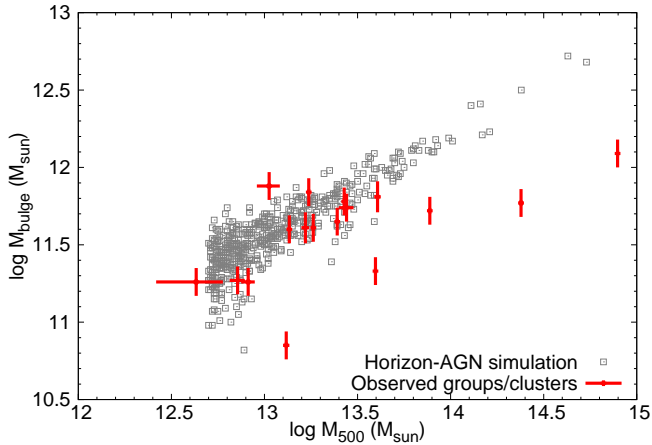}
       \vspace{0cm}
      \caption{The relation between the stellar mass and the $M_{\rm 500}$ mass for the 17 galaxy groups/clusters studied in this work and for the systems in the Horizon-AGN simulation. Note that the observed relation is notably flatter than the simulated one}
     \label{fig:fig2}
  \end{center}
\end{figure}

\subsection{Comparing the observed and simulated relations}

We compare data and simulation in Figures~\ref{fig:fig1} and \ref{fig:fig2}. Simulation and data agree remarkably well in the $M_{\rm BH}-M_{\rm 500}$ plane for the entire range of masses analyzed here (Figure~\ref{fig:fig1} left panel). The scatter is also similar, albeit slightly lower ($\sigma_{\rm X} = 0.20$ and $\sigma_{\rm Y} = 0.19$) than that obtained for the observed sample. 

The   $M_{\rm BH}-M_{\rm bulge}$ relation, instead, shows several differences  (Figure~\ref{fig:fig1} right panel). The simulated galaxies have stellar masses reaching $5\times10^{12} \ \rm{M_{\odot}}$, while observations are limited to $10^{12} \ \rm{M_{\odot}}$. Additionally, the relation for simulated galaxies is much tighter. The tightness of the relation between BH mass and galaxy mass was already noticed in \cite{volonteri2016}, who argued that, although merger histories and environmental effects induce some scatter, unresolved internal process represent an important contribution to capturing the full scatter in the relationships. 

For completeness, as in the case of observations, we compare $M_{\rm bulge}$ and $M_{\rm 500}$ (Figure \ref{fig:fig2}). Observations and the simulation agree well at $M_{\rm 500}<3\times10^{13} \ \rm{M_{\odot}}$, but in larger groups and in clusters, the stellar mass of simulated galaxies keeps rising steadily, while, as noted already for the   $M_{\rm BH}-M_{\rm bulge}$ relation, in observations, galaxy mass seems to saturate at  $\sim 10^{12} \ \rm{M_{\odot}}$. 

 \begin{figure}[!]
  \begin{center}
    \leavevmode
      \epsfxsize=0.48\textwidth\epsfbox{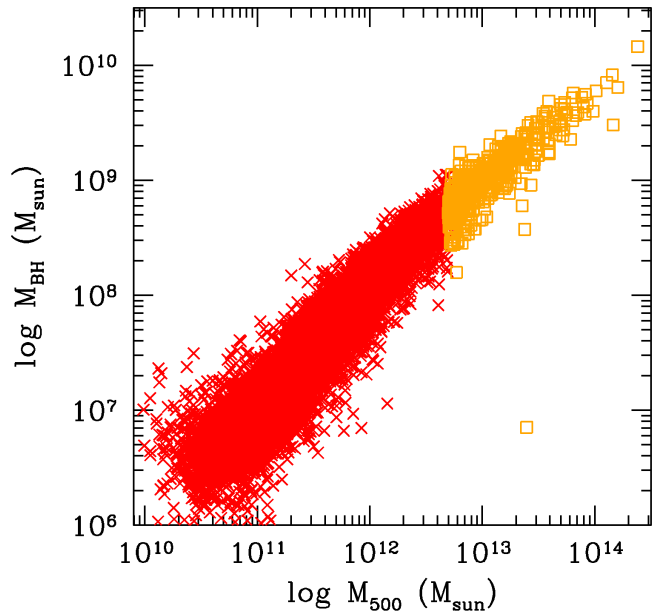}
       \vspace{0cm}
      \caption{The relation between the stellar mass and the halo mass, in the Horizon-AGN simulation for all central galaxies in halos. BGGs/BCGs are shown as squares. The relation is shallower and tighter at the high-mass end (see Table~\ref{tab:fit_sim}).}
     \label{fig:fig3}
  \end{center}
\end{figure}

\begin{table}
\caption{Best-fit parameters on linear regression for $M_{\rm BH}-M_{\rm 500}$ in the simulation}
\begin{minipage}{8.5cm}
\renewcommand{\arraystretch}{1.5}
\centering
\begin{tabular}{c c c c c c c}
\hline 
 &  $\alpha$  & $\beta$ & $\sigma$  \\
\hline
All halos 					   & -1.23 & 0.77   & 0.42 \\
$M_{\rm 500}<2.5\times10^{12}$ & -0.69 &  0.72  & 0.42 \\
$M_{\rm 500}>2.5\times10^{12}$ & -1.22 &  0.78 & 0.18 \\
 \hline \\
\end{tabular} 
\end{minipage}
\vspace{0.5cm}
\label{tab:fit_sim}
\end{table}  

When comparing with simulations, we find that the Horizon-AGN simulation reproduces very well the observations in range explored here. If we extend the analysis to lower halo masses, considering only main halos and not satellites \citep[as satellites are affected by tidal stripping, see][]{volonteri2016}, we find that the relation steepens slightly and scatter increases (Figure \ref{fig:fig3}). If we fit separately the low- and high-mass ends with a divide at $M_{\rm 500}=2.5\times10^{12}$, we find $\sigma=0.42$ and $\sigma=0.18$ respectively (see Table~\ref{tab:fit_sim}). As noted above for the relation with bulge mass, the Horizon-AGN simulation is likely to underestimate scatter at the low-mass end; therefore, the scatter found here is a lower limit. The limited correlation in late-type and bulgeless  galaxies between BHs and dark matter halos \citep{kormendy11,sun13} may be an illustration of this effect.

\subsection{Tight relation between the BH mass and total halo mass}

Our conclusions, namely that the BH mass of BGGs/BCGs and the total mass of galaxy groups and clusters exhibit a tight relation, is similar to that obtained through the study of globular clusters. Specifically, \citet{blakeslee97} suggested that the specific frequency of globular clusters in BCGs tightly correlates with the total mass of the host galaxy clusters. More recently, this conclusion was expanded for all types of galaxies: \citet{hudson14} established that the mass ratio between globular clusters and the total halo is a constant ratio for every mass range. In addition, \citet{burkert10} demonstrated that the specific frequency of globular clusters tightly correlate with the BH mass of galaxies. Combining these results implies that the mass of BHs are closely connected to the total halo mass of galaxies and for BCGs the BH mass and the total cluster mass are related. 

In our sample of BGGs/BCGs, the BH mass appears to trace halo mass, via gas temperature, more tightly than the bulge mass. This suggests that the growth of the BHs in BGGs/BCGs may be indirectly governed by physical processes that are influenced by the large-scale potential of the host galaxy group or clusters. Such processes may be due to the inflow of cold gas onto the cluster center that directly feeds the BH, a series of low angular momentum mergers, or residual gas in the center of the galaxy after most of the gas was heated/expelled due to the quasar activity. To differentiate between these possibilities and to constrain the origin of the tight $M_{\rm BH} - M_{\rm 500}$ relation, it will be essential to explore the evolutionary history of both the BGGs/BCGs and their host galaxy groups/clusters. However, such a dedicated study is beyond the scope of the present paper.

\section{Conclusions}

In this work, we explored the correlation between the BH mass of BGGs/BCGs and the properties of the host galaxy and group/cluster. Our results can be summarized as follows. 

\begin{enumerate}

\item We analyzed \textit{XMM-Newton} observations of 17 galaxy groups and clusters and measured their best-fit gas temperatures. The temperatures range from $0.347-8.165$ keV, which corresponds to $M_{\rm 500} = (0.43-78.90)\times10^{13} \ \rm{M_{\odot}}$. 
\item By utilizing dynamical BH mass measurements and the best-fit gas temperatures of the groups/clusters, we establish the $M_{BH} - kT$ relation, which exhibits a tight correlation with Pearson and Spearman correlation coefficients of $0.97$ and $0.83$, respectively.
\item We established the  $M_{\rm BH} - M_{\rm bulge} $ relation, which exhibits large scatter and a Pearson and Spearman correlation coefficients of $0.70$ and $0.35$, respectively. 
\item We conclude that for BHs residing in the centers galaxy groups and clusters, the scaling relation with the large-scale halo properties are significantly tighter than with the stellar bulge mass of the host galaxy. This suggests that the BH mass of BGGs/BCGs may be set by physical processes that are governed by the properties of the host galaxy cluster. 
\item We compare the observed relations with those obtained in the Horizon-AGN simulation. We find that the overall slope of the relations are similar, albeit the simulated $M_{\rm BH} - M_{\rm bulge} $ exhibits significantly smaller scatter than the observed one. \\

\end{enumerate}

\bigskip 

\begin{small}
\noindent
\textit{Acknowledgements.}
We thank the referee for the careful review. This work uses observations obtained with \textit{XMM-Newton}, an ESA science mission with instruments and contributions directly funded by ESA Member States and NASA. In this work, the NASA/IPAC Extragalactic Database (NED) have been used. \'A.B. acknowledges support from the Smithsonian Institution. M.V. acknowledges funding from the European Research Council under the European CommunityÕs Seventh Framework Programme (FP7/2007-2013 Grant Agreement no. 614199, project ÒBLACKÓ). 
\end{small}

\end{document}